\documentclass[11pt]{article}
\usepackage{graphicx}
\usepackage{amsmath}
\usepackage{amssymb}
\usepackage{amsxtra}

\title{Using Two Measures Theory to Approach Bags and Confinement}
\author{E. I. Guendelman\thanks{e-mail: guendel@bgu.ac.il}
Physics Department, Ben Gurion University of the Negev 
\\Beer Sheva 84105,
Israel}

\begin{document}
\maketitle

\begin{abstract}
We consider the question of bags and confinement in the framework of a theory which uses two volume elements  $\sqrt{-{g}}d^{4}x$ and $\Phi d^{4}x$, where $\Phi $ is a metric independent density. For scale invariance a dilaton field $\phi$  is considered. 
Using the first order formalism, curvature ( $\Phi R$ and  $\sqrt{-g}R^{2}$ ) terms , gauge field term( $\Phi\sqrt { - F_{\mu \nu }^{a}\, F^{a}_{\alpha\beta}g^{\mu\alpha}g^{\nu\beta}}$ and  $\sqrt{-g} F_{\mu \nu }^{a}\, F^{a}_{\alpha\beta}g^{\mu\alpha}g^{\nu\beta}$ ) and dilaton kinetic terms are introduced in a conformally invariant way. Exponential potentials for the dilaton break down (softly) the conformal invariance down to global scale invariance, which also suffers s.s.b. after integrating the equations of motion. The model has a well defined flat space limit.  As a result of the s.s.b. of scale invariance phases with different vacuum energy density appear. Inside the bags the gauge dynamics is normal, that is non confining, while for the outside, the gauge field dynamics is confining.  
\end{abstract}

\section{Introduction and Conclusions}\label{eg1s:intro}
In the bag model of confinement \cite{eg1MIT} two phases for gauge fields are identified, first the free non confining dynamics for 
the gauge fields that holds inside the bags, there gauge fields are prevented to flow to the outside (confinement) region
by the M.I.T. bag model boundary conditions. 
On the other hand, in modern cosmology working with different phases  is a central theme and also in the context of modern cosmology, as well as for the bag model, we need two phases. In cosmology the two phases they should be connected
through cosmological evolution, while in the bag model through the boundary of the bag.

As it is well known, in the context of cosmology, it is very difficult to understand the smallness of the observed present vacuum energy density.
This "cosmological constant problem", has been reformulated in the context of the two measures theory (TMT) \cite{eg1TMT1} - \cite{eg1TMT5}
and more specifically in the context of the scale invariant
realization of such theories  \cite{eg1TMT2} - \cite{eg1TMT5}. These theories
can provide a new approach to  the cosmological constant problem and can be generalized to obtain also
a theory with a dynamical space-time \cite{eg1dyn} . The TMT models consider two measures of integration in the action,
the standard $\sqrt{-g}$ where $g$ is the determinant of the metric and another measure $\Phi$ independent
of the metric. To implement scale invariance (S.I.), a dilaton  field is introduced \cite{eg1TMT2} - \cite{eg1TMT5}.
  
In the TMT theories we obtain drastic modifications of the dynamics of vacuum energy density, which produces naturally a 
zero cosmological constant and together with this regions of very small vacuum energy density. These ideas work particularly well in the context of scale invariance which can be spontaneously broken by the integration of the equations of motion. What is most important for the present research is that it is the nature of the two measures theories to change not only the dynamics of the vacuum energy density, but also that of the matter itself. For example, in the context of the spontaneously broken scale invariant theories, the dilaton field decouples from the fermionic matter at high densities, avoiding the fifth force problem, see \cite{eg1TMT6}, \cite{eg1TMT7}. On the opposite limit, fermionic matter was shown to contribute to the dark energy density for very small densities \cite{eg1TMT8}.

In this paper our focus will be on the interplay between gauge field dynamics, in particular confinement properties and the different phases as defined with the help  of TMT and whether the possibility of obtaining a confinement phase and a deconfined phase (like in the MIT bag model) can addressed in this context.
 Using the first order formalism, curvature ( $\Phi R$ and  $\sqrt{-g}R^{2}$ ) terms , gauge field terms and dilaton kinetic 
  terms can be introduced in a conformally invariant way. Exponential potentials for the dilaton break down (softly) the conformal invariance down to global scale invariance, which also suffers s.s.b. after integrating the equations of motion. As a result of the s.s.b. of scale invariance phases with different vacuum energy density appear. 
In this contribution we will review the principles
of the TMT and in particular the model studied in \cite{eg1TMT2}, which has
global scale invariance. Then, we
look at the generalization of this model \cite{eg1TMT5} by adding a curvature square or simply "$R^{2}$ term"
and show that the resulting model contains now two flat
regions. The existence of two flat regions for the potential
is shown to be consequence of the s.s.b. of the scale symmetry. The model is then further extended to include gauge fields. A gauge field strength squared term coupled to $\sqrt{-g}$,  a square root of a gauge field strength squared term coupled to $\Phi$ and a mass term for the gauge fields coupled to $\Phi$ are the unique candidates which respect local conformal invariance and they can provide a consistent framework to answer the questions posed. For the issue of electric confinement we disregard the mass term and consider only the gauge field strength squared term coupled to $\sqrt{-g}$ and  the square root of a gauge field strength squared term coupled to $\Phi$. This square root term has been studied before in order to reproduce confinement behavior 
\cite{eg1Guendelman},\cite{eg1Guendelman-Gaete}-\cite{eg1Korover3}. In the context of the softly broken conformally invariant TMT model it appears however in a particularly natural way. After s.s.b. of scale invariance, the amazing feature that the square root gauge field term is totally screened in the high vacuum energy regions (inside the bags) and acts only outside the bags, reproducing basic qualitative behavior postulated in the M.I.T bag model, also some difficulties present in the original formulation of the square root gauge fields approach to confinement are resolved when the square root term is embedded in the TMT model presented here\ldots.

\section{The Two Measures Theory Fundamentals}

We work in the context of a theory built along the lines of the two measures theory (TMT) \cite{eg1TMT1}, \cite{eg1TMT2}, \cite{eg1TMT3}, which deals with actions of the form,
\begin{equation}\label{eg1e6}
S = \int L_{1} \sqrt{-g}d^{4}x + \int L_{2} \Phi  d^{4} x    
\end{equation}
where $\Phi$ is an alternative "measure of integration", a density independent of the metric, for example in terms of four scalars $\varphi_{a}$ ($a = 1,2,3,4$),it can be  obtained as follows:

\begin{equation}\label{eg12}
\Phi =  \varepsilon^{\mu\nu\alpha\beta}  \varepsilon_{abcd}
\partial_{\mu} \varphi_{a} \partial_{\nu} \varphi_{b} \partial_{\alpha}
\varphi_{c} \partial_{\beta} \varphi_{d}
\end{equation}
and more specifically work in the context of the globally scale invariant
realization of such theories  \cite{eg1TMT2}, \cite{eg1TMT3}, which require the introduction of a dilaton field $\phi$.
We look at the generalization of these models \cite{eg1TMT3} where an "$R^{2}$ term" is present, 
\begin{equation}\label{eg1e10}
L_{1} = U(\phi) + \epsilon R(\Gamma, g)^{2} 
\end{equation}
\begin{equation}\label{eg1e11}
L_{2} = \frac{-1}{\kappa} R(\Gamma, g) + \frac{1}{2} g^{\mu\nu}
\partial_{\mu} \phi \partial_{\nu} \phi - V(\phi)
\end{equation}
\begin{equation}\label{eg1e12}
R(\Gamma,g) =  g^{\mu\nu}  R_{\mu\nu} (\Gamma) , R_{\mu\nu}
(\Gamma) = R^{\lambda}_{\mu\nu\lambda}
\end{equation}
\begin{equation}\label{eg1e13}
R^{\lambda}_{\mu\nu\sigma} (\Gamma) = \Gamma^{\lambda}_
{\mu\nu,\sigma} - \Gamma^{\lambda}_{\mu\sigma,\nu} +
\Gamma^{\lambda}_{\alpha\sigma}  \Gamma^{\alpha}_{\mu\nu} -
\Gamma^{\lambda}_{\alpha\nu} \Gamma^{\alpha}_{\mu\sigma}.
\end{equation}
For the case the potential terms
$U=V=0$ we have local conformal invariance

\begin{equation}\label{eg1e14}
g_{\mu\nu}  \rightarrow   \Omega(x)  g_{\mu\nu}
\end{equation}
and $\varphi_{a}$ is transformed according to
\begin{equation}\label{eg1e15}
\varphi_{a}   \rightarrow   \varphi^{\prime}_{a} = \varphi^{\prime}_{a}(\varphi_{b})
\end{equation}

\begin{equation}\label{eg1e16}
\Phi \rightarrow \Phi^{\prime} = J(x) \Phi    
\end{equation}
where $J(x)$  is the Jacobian of the transformation of the $\varphi_{a}$ fields.
This will be a symmetry in the case $U=V=0$ if 
\begin{equation}\label{eg1e17}
\Omega = J
\end{equation}

global scale invariance is satisfied if \cite{eg1TMT3}, \cite{eg1TMT2}($ f_{1}, f_{2},\alpha $ being constants),
\begin{equation}\label{eg1e19} 
V(\phi) = f_{1}  e^{\alpha\phi},  U(\phi) =  f_{2}
e^{2\alpha\phi}
\end{equation}

 In the variational principle $\Gamma^{\lambda}_{\mu\nu},
g_{\mu\nu}$, the measure fields scalars
$\varphi_{a}$ and the "matter" - scalar field $\phi$ are all to be treated
as independent
variables although the variational principle may result in equations that
allow us to solve some of these variables in terms of others, that is, the first order formalism is employed, where any relation between the connection coefficients and the metric is obtained from the variational principle, not postulated a priori. A particularly interesting equation is the one that arises from the $\varphi_{a}$
fields, this yields $L_2 = M$, where $M$ is a constant that spontaneously breaks scale invariance.
The 
Einstein frame,  which is a redefinition of the metric by a conformal factor, is defined as 

\begin{equation}\label{eg1e47}
\overline{g}_{\mu\nu} = (\chi -2\kappa \epsilon R) g_{\mu\nu}
\end{equation}
where $\chi$ is the ratio between the two measures, $\chi =\frac{\Phi}{\sqrt{-g}}$, is determined from the consistency of the equations to be $\chi = \frac{2U(\phi)}{M+V(\phi)}$. The relevant fact is that the connection coefficient equals the Christoffel symbol of this new metric (for the original metric this "Riemannian" relation does not hold). There is an effective action, where one can use this Einstein frame metric, it is determined by the pressure functional, 
($X = \frac{1}{2} \overline{g}^{\mu\nu}\partial_{\mu} \phi \partial_{\nu} \phi $).

\begin{equation}
S_{eff}=\int\sqrt{-\overline{g}}d^{4}x\left[-\frac{1}{\kappa}\overline{R}(\overline{g})
+p\left(\phi,R\right)\right] \label{eg1k-eff}
\end{equation}

\begin{equation}
 p = \frac{\chi}{\chi -2 \kappa \epsilon R}X - V_{eff}
\end{equation}
where $V_{eff}$ is an effective potential for the dilaton field given by

\begin{equation}\label{eg1e50}
 V_{eff}  = \frac{\epsilon R^{2} + U}{(\chi -2 \kappa \epsilon R)^{2} }
\end{equation}
$\overline{R}$ is the Riemannian curvature scalar built out of the bar metric, $R$ on the other hand is the non Riemannian curvature scalar defined in terms of the connection and the original metric,which turns out to be given by $R = \frac{-\kappa (V+M) +\frac{\kappa}{2} \overline{g}^{\mu\nu}\partial_{\mu} \phi \partial_{\nu} \phi \chi}
{1 + \kappa ^{2} \epsilon \overline{g}^{\mu\nu}\partial_{\mu} \phi \partial_{\nu} \phi}$. This $R$ can be inserted in the action (\ref{eg1k-eff})
or alternatively, $R$ in the action (\ref{eg1k-eff}) can be treated as an independent degree of freedom, then its variation gives the required value
as one can check (which can then be reinserted in (\ref{eg1k-eff})).
Introducing this $R$ into the expression (\ref{eg1e50}) and considering a constant field $\phi$ we find that $ V_{eff}$ has two flat
regions. The existence of two flat regions for the potential
is shown to be consequence of the s.s.b. of the scale symmetry (that is of considering  $M \neq 0$ ).  

\section{ Gauge Field Kinetic Terms, Mass Terms and "Confinement Terms" in the Softly Broken Conformally Invariant TMT Model }

Now we will see that the incorporation of a term of the form 
$ \sqrt { - F_{\mu \nu }^{a}\, F^{a\mu \nu } }$, which in flat space is known to introduce confinement behavior, is
in the TMT case  quite natural, in fact, there is a good reason to include it, since it respects conformal symmetry if coupled to the new measure $\Phi$ . This kind of coupling of a square root gauge field strength to a new measure has been considered in the context of conformally invariant braneworld scenarios\cite{eg1sqrt}-\cite{eg1sqrt3}, which allow compactification, branes and zero four dimensional cosmological constant.
Another place where square root of gauge field square coupled to a  modified measure find a natural place is in the formulation of Weyl invariant brane theories\cite{eg1sqrtbwill}-\cite{eg1sqrtbwill3}. Black hole solutions in the presence of both a regular Maxwell term and a square root of gauge field square have been also considered \cite{eg1blackholes}.
An early model which enriches the "square root" gauge theory with a dilaton field so that it could describe confined and
unconfined regions  (bags) was done "by hand" in \cite{eg1confdeconf}. This will be obtained most elegantly however by embedding the square root terms into the TMT  formalism.

The reason for the conformal invariance of the $ \sqrt { - F_{\mu \nu }^{a}\, F^{a\mu \nu } }$ is very simple: conformally invariant terms (with respect to (\ref{eg1e14}) , (\ref{eg1e15}), (\ref{eg1e16}) and (\ref{eg1e17}))in TMT  are of two kinds, if they multiply the measure $\Phi$ they they must have homogeneity 1 with respect to $g^{\mu \nu}$, or if they multiply the measure $\sqrt{-g}$ they they must have homogeneity 2   with respect to $g^{\mu \nu}$, since 
$\sqrt { - F_{\mu \nu }^{a}\, F^{a\mu \nu }}$ $=\sqrt { - F_{\mu \nu }^{a}\, F^{a}_{\alpha\beta}g^{\mu\alpha}g^{\nu\beta}}$, then according to (\ref{eg1e14}) $\sqrt { - F_{\mu \nu }^{a}\, F^{a\mu \nu }}\rightarrow \Omega^{-1}\sqrt { - F_{\mu \nu }^{a}\, F^{a\mu \nu }}$ if $\Omega>0$
and $\Phi \rightarrow  J \Phi= \Omega \Phi$, so that $\Phi\sqrt { - F_{\mu \nu }^{a}\, F^{a\mu \nu }}$ is invariant, provided $J=\Omega>0$.

A similar story happens with a mass term for the gluon, $A_{\mu }^{a}\, A^{a}_{\alpha}g^{\mu\alpha}$ in TMT, this can be a conformally invariant
if it goes multiplied with the measure $\Phi$.

Likewise, the  conformally invariance implies that a term proportional to
 $F_{\mu \nu }^{a}F^{a}_{\alpha\beta}g^{\mu\alpha}g^{\nu\beta}$ has come multiplied by 
the Riemannian measure $\sqrt{-g}$, since $\sqrt{-g}F_{\mu \nu }^{a}F^{a}_{\alpha\beta}g^{\mu\alpha}g^{\nu\beta}$ is invariant under conformal transformations of the metric. We take therefore for our softly broken conformal invariant model, where we exclude mass terms for the gluons,

\begin{equation}\label{eg1finalaction}
S = S_L + S_{R^2} - \frac  {1}{4}\int d^4x\sqrt{-g}F_{\mu \nu }^{a}F^{a}_{\alpha\beta}g^{\mu\alpha}g^{\nu\beta} 
+\frac{N}{2}\int d^4x \Phi\sqrt{-F_{\mu \nu }^{a}F^{a}_{\alpha\beta}g^{\mu\alpha}g^{\nu\beta}}
\end{equation}
here $S_L$ contains the terms linear in the curvature scalar, plus scalar field kinetic terms and potentials and $S_{R^2}$ refers to the $R^2$ contribution defined before. The consequences of having both a mass term and a confinement term have been explored in \cite{eg1interplay} where it was shown that in such a case confinement is lost in favor of a Coulomb like behavior, but, as mentioned before, for the purposes of this paper this will not be considered here.

\section{Description of the Bag Dynamics in the Softly Broken Conformally Invariant TMT Model }
Let us proceed now to describe the consequences of the action (\ref{eg1finalaction}). 
The steps to follow are the same as in the case where we did not have gauge fields.

One interesting fact is that the terms that enter the constraint that determines $\chi$ are only those that break the
conformal invariance and  the constant of integration $M$.  Since the new terms involving the gauge fields do not break the conformal invariance 
(\ref{eg1e14}), (\ref{eg1e15}), (\ref{eg1e16}), (\ref{eg1e17}), the relevant terms that violate this symmetry are only the $U$ and $V$ terms and the constraint remains the same. We can then continue and construct all the equations of motion as before.

The easiest way to summarize the result of such analysis is to consider the effective action in the Einstein frame, as we did in the previous case where we did not have gauge fields.
Now, for the case where gauge fields are included in the way described by (\ref{eg1finalaction}), 
all the equations of motion in the Einstein frame will be correctly described by 
\begin{equation}\label{eg1finalbagmodel}
S_{eff}=\int\sqrt{-\bar{g}}d^{4}x\left[-\frac{1}{\kappa}\bar{R}(\bar{g})
+p\left(\phi,R,X,F_{\mu \nu }^{a}, \bar{g}^{\alpha\beta}\right)\right] 
\end{equation}

\begin{equation}\label{eg1newp}
 p = \frac{\chi}{\chi -2 \kappa \epsilon R} \left[X
+\frac{N}{2}\sqrt{-F_{\mu \nu }^{a}F^{a}_{\alpha\beta}\bar{g}^{\mu\alpha}\bar{g}^{\nu\beta}}\right] - \frac{1}{4}F_{\mu \nu }^{a}F^{a}_{\alpha\beta}\bar{g}^{\mu\alpha}\bar{g}^{\nu\beta} 
 - V_{eff}
\end{equation}

\begin{equation}\label{eg1newV}
 V_{eff}  = \frac{\epsilon R^{2} + U}{(\chi -2 \kappa \epsilon R)^{2} }
\end{equation}
where 
\begin{equation}\label{eg1chi3}
\chi = \frac{2U(\phi)}{M+V(\phi)}
\end{equation}
We have again two possible formulations concerning $R$:
Notice first that $\bar{R}$ and $R$ are different objects, the $\bar{R}$ is the Riemannian curvature scalar in the Einstein frame,
while $R$ is a different object. This $R$ will be treated in two different ways:

1. First order formalism for $R$. Here $R$ is a lagrangian variable, determined as follows,  $R$ that appear in the expression above for $p$ can be obtained from the variation of the pressure functional action above with respect to $R$, this gives exactly the expression for $R$ that can be solved for by using the equations of motion in the original frame (and then reexpresing the result in terms of the bar metric), in terms of $X, \phi$, etc. 

2. Second order formalism for $R$. $R$ that appear in the action above is exactly the expression for $R$ which can be solved from the equations of motion in terms of $X, \phi$, etc. Once again, the second order formalism can be obtained from the first order formalism by solving algebraically R from the eq. obtained by variation of $R$ , and inserting back into the action.
Now $R$ is given by
\begin{equation}\label{eg1newR}
R = \frac{-\kappa (V+M) +\kappa \chi \left( X +\frac{N}{2}\sqrt{-F_{\mu \nu }^{a}F^{a}_{\alpha\beta}\bar{g}^{\mu\alpha}\bar{g}^{\nu\beta}}\right)}
{1 + 2\kappa ^{2} \epsilon \left(X +\frac{N}{2}\sqrt{-F_{\mu \nu }^{a}F^{a}_{\alpha\beta}\bar{g}^{\mu\alpha}\bar{g}^{\nu\beta}}\right) }
\end{equation}

\section{Regular gauge field dynamics inside the bags}
From (\ref{eg1finalbagmodel}), (\ref{eg1newp}) and (\ref{eg1chi3}), we see that the $N$ term, responsible for the confining gauge dynamics, gets dressed in the Einstein frame effective action by the factor $\frac{\chi}{\chi -2 \kappa \epsilon R}$, we will have to check also whether $V_{eff}$  contributes to the gauge field equations of motion.

As we consider regions inside the bags, where $\phi \rightarrow -\infty$, we see  that $\chi$ as given by (\ref{eg1chi3}), approaches zero for 
$M \neq 0$, for the case therefore $\epsilon \neq 0$
the $N$ term inside the bags disappears. Notice that if we had not introduced the curvature squared term (i.e. if $\epsilon = 0$) this effect would be absent.

In this same limit and with the same conditions, using only that as $\phi \rightarrow -\infty$, $U \rightarrow 0$ and $\chi \rightarrow 0 $, we see that still, in the more complicated theory with gauge fields the same bag constant  $V_{eff} \rightarrow  \frac{1}{4\epsilon \kappa^{2}}$ is obtained, so $V_{eff}$ does not contribute to the gauge field equations of motion, but does provide the Bag constant.

 In the limit $\phi \rightarrow -\infty$, the only term providing gauge field dynamics is the standard term 
$-\frac{1}{4}F_{\mu \nu }^{a}F^{a}_{\alpha\beta}\bar{g}^{\mu\alpha}\bar{g}^{\nu\beta} $.

\section{Confining gauge field effective action outside the bags}
 We are going to assume $M>0$, so to keep $\chi$ positive and finite everywhere and take now the opposite limit, $\phi \rightarrow +\infty$ .
 Furthermore, the choice $M>0$ pushes the scalar field outside the bag to large values of $\phi$, since the absolute minimum of the effective potential is found for such values, then   confining dynamics appears,

\begin{equation}\label{eg1Vlimit}
 V_{eff}   \rightarrow C+4B \left[\,X
+\frac{N}{2}\sqrt{-F_{\mu \nu }^{a}F^{a}_{\alpha\beta}\bar{g}^{\mu\alpha}\bar{g}^{\nu\beta}}\ \right]^2
\end{equation}

and
 
 $ \frac{\chi}{\chi -2 \kappa \epsilon R}\left[\,X
+\frac{N}{2}\sqrt{-F_{\mu \nu }^{a}F^{a}_{\alpha\beta}\bar{g}^{\mu\alpha}\bar{g}^{\nu\beta}}\ \right]  \rightarrow $ 

\begin{equation}
A\left[\,1+2\kappa^2 \epsilon \left[\, X
+\frac{N}{2}\sqrt{-F_{\mu \nu }^{a}F^{a}_{\alpha\beta}\bar{g}^{\mu\alpha}\bar{g}^{\nu\beta}}\ \right]\ \right] \left[\,X
+\frac{N}{2}\sqrt{-F_{\mu \nu }^{a}F^{a}_{\alpha\beta}\bar{g}^{\mu\alpha}\bar{g}^{\nu\beta}}\ \right]
\end{equation}
where the constants $A$, $B$ and $C$ are given by, $A = \frac{f_2}{f_2 + \kappa^2\epsilon f_1^2}$, $B = \frac{\epsilon \kappa^2}{4}A$ and
$C   =\frac{f_1^2}{4f_2}\,A$.
Therefore, the resulting dynamics outside the bag, for $\phi \rightarrow +\infty$ will be described by the effective action (expressing $B$ in terms of $A$),

\begin{equation}\label{eg1outsidebag}
S_{eff, out}=\int\sqrt{-\bar{g}}d^{4}x\left[-\frac{1}{\kappa}\bar{R}(\bar{g})
+p_{out}\left(\phi,X, F\right)\right] 
\end{equation}

\[ \nonumber
p_{out}\left(\phi,X, F\right) = AX+ A \frac{N}{2}\sqrt{-F_{\mu \nu }^{a}F^{a}_{\alpha\beta}\bar{g}^{\mu\alpha}\bar{g}^{\nu\beta}} 
- (1 + N^2\epsilon \kappa^2A)\frac{1}{4}F_{\mu \nu }^{a}F^{a}_{\alpha\beta}\bar{g}^{\mu\alpha}\bar{g}^{\nu\beta}
\]
\begin{equation}\label{eg1poutsidebag2}
+ AN\epsilon \kappa^2 X\sqrt{-F_{\mu \nu }^{a}F^{a}_{\alpha\beta}\bar{g}^{\mu\alpha}\bar{g}^{\nu\beta}} + A\epsilon \kappa^2X^{2} -C
\end{equation}

Full details concerning these developments have been presented elsewhere \cite{eg1GuendelmanIJMPA}. Working in the case  where gravitation plays an important role, one could also think of using the approach developed here to generalize the "hiding" \cite{eg1hide}
and "hiding and confining effects"\cite{eg1hide-conf}, where the confining region is an uncompactified space-time and where charges send the gauge field flux they generate completely into a "flux tube-like" compactified region.

\section*{Acknowledgements}
It is a pleasure for me to thank Norma Mankoc\v c, Holger Nielsen and all the organizers of the Bled workshop "On What Comes Beyond The Standard Models?" for a very productive and  interesting conference.

\end{document}